\documentstyle[epsf,prc,aps,graphicx]{revtex}
\begin{document}
\preprint{Saclay-T00/049}
\draft
\title{Is the analysis of flow at the CERN SPS reliable?}
\author{Nicolas BORGHINI$^1$, Phuong Mai DINH$^2$, and Jean-Yves OLLITRAULT$^2$}
\address{$^1$~Laboratoire de Physique Th\'eorique des Particules El\'ementaires 
\\ 
Universit\'e Pierre et Marie Curie\\
4, place Jussieu\\
F-75252 Paris cedex 05}
\address{$^2$~Service de Physique Th\'eorique, CEA-Saclay\\
F-91191 Gif-sur-Yvette cedex}
\maketitle
\begin{abstract}
Several heavy ion experiments at SPS have measured azimuthal 
distributions of particles with respect to the reaction plane. 
These distributions are deduced from two-particle azimuthal 
correlations under the assumption that they result solely from 
correlations with the reaction plane. 
In this paper, we investigate other sources of azimuthal correlations:
transverse momentum conservation, which produces back-to-back 
correlations, resonance decays such as $\Delta\to p\pi$, 
HBT correlations and final state interactions. 
These correlations increase with impact parameter: most of 
them vary with the multiplicity $N$ like $1/N$. 
When they are taken into account, 
the experimental results of the NA49 collaboration at SPS 
are significantly modified. 
These correlations might also explain an important fraction of 
the pion directed flow observed by WA98.
Data should be reanalyzed taking into account carefully 
these non--flow correlations. 
\end{abstract}

\section{Introduction}

In ultrarelativistic heavy ion collisions, azimuthal 
distributions of the produced particles with respect to the 
reaction plane are of particular interest. Their 
anisotropies, which we refer to as ``flow'', result from the 
interactions between the produced particles. 
They thus probe the hot stages of the collision and may provide 
valuable information on the state of the interaction region: 
thermalized or not, equation of state, 
etc... \cite{Olli92,DANI4,Henning,Zhang}
In the recent years, flow has been measured 
by many experiments at the Brookhaven AGS
with Au projectiles \cite{E877,E895,E917} and at the CERN SPS 
with Pb projectiles \cite{Data,WA98,WA98bis,NA45,NA52,NA50}.
Since the orientation of the reaction plane is not known 
a priori, flow measurements are usually extracted from 
two-particle azimuthal correlations. 
This is based on the idea that azimuthal correlations between 
two particles are generated by the correlation of the azimuth 
of each particle with the reaction plane. 

The assumption that this is the only source 
of two particle azimuthal correlations 
dates back to the early days of the flow 
\cite{DANI}. 
It still underlies the analyses done at 
ultrarelativistic energies, both at AGS 
and at SPS (see, however, \cite{WA93}). 
However, there are various sources of direct 
azimuthal correlations between particles, which do not 
involve the reaction plane. 
We have shown in a recent paper~\cite{nous} 
that the quantum correlations due to the HBT effect 
yield azimuthal correlations. When they are taken into 
account, the results obtained for the charged pion flow 
by the NA49 collaboration~\cite{Data} change significantly.
Here, we investigate other sources of non--flow azimuthal correlations. 
A first source is the condition 
that the total transverse momentum of the outgoing 
particles is zero: this gives a back-to-back correlation between 
their momenta, which can be evaluated quantitatively. 
This effect is usually taken into account at lower 
energies \cite{DANI2} but has been neglected so far at AGS and SPS. 
Other correlations are due to resonance decays such as 
$\Delta\to p\pi$ and $\rho\to\pi\pi$. 
Finally, Coulomb and strong interactions between pairs of particles
with low relative velocities (which we refer to as 
``final state interactions'') produce small angle azimuthal 
correlations\cite{FSI}. 

Most of these correlations are typically of order $1/N$, 
where $N$ is the number of particles emitted in
the collision. At SPS energies, correlations due to flow are
so small that additional correlations become of the same 
order, and cannot be neglected. 
The $1/N$ dependence also determines the variation of 
non--flow correlations with the centrality of the collision. 
Therefore they increase with impact parameter 
up to very peripheral collisions; on the other hand,
they do not vanish for 
central collisions where the flow is zero by symmetry. 
They are very important at large impact parameters and 
should be taken into account, in particular, when 
studying the centrality dependence of the flow, which has 
been recently proposed as a sensitive probe of the phase transition 
to the quark--gluon plasma \cite{Sorge,Volo99,Bravina}. 

In section \ref{s:sec2}, we briefly recall 
how the flow can be extracted from two--particle azimuthal
correlations. 
We use the same method and notations as in \cite{nous}.
The various sources of non--flow correlations are reviewed and 
their effects are estimated in section \ref{s:sources}. 
The experimental data obtained by the NA49 and WA98 collaborations 
for Pb--Pb collisions at SPS are discussed in section \ref{s:spsdata}. 
We show in particular that the effect of momentum conservation 
alone is large enough to reverse the sign of the proton directed
flow measured by NA49~\cite{Data}, 
which is used to define the reaction plane. 
$\Delta$ decays produce correlations which are of the opposite sign,
and of the same order of magnitude, although we do not attempt to 
calculate them accurately. 
For WA98, final state interactions play an important part, and 
may explain a large fraction of the observed pion directed flow. 
Our conclusions are given in section \ref{s:sec5}.

\section{Two--particle azimuthal correlations and standard 
flow analysis}
\label{s:sec2}

In this section, we define the flow and non--flow contributions to 
two--particle
azimuthal correlations, and we show how azimuthal distributions 
with respect to the reaction plane can be extracted from these
correlations. 

We call ``flow'' the azimuthal correlations between 
the outgoing particles and the reaction plane. 
These are best conveniently characterized in terms of the Fourier 
coefficients $v_n$~\cite{VOLO/ZH} which we now define. 
We choose a coordinate system in which the $x$ axis is the impact 
direction, and $(x,z)$ the reaction plane. 
$\phi$ denotes the azimuthal angle with respect to the
reaction plane.
In this frame, $v_n$ can be expressed as a
function of the one-particle momentum distribution 
$dN_j/d^3{\bf p}$ for a particle of type $j$ 
(in this paper,  we consider pions and protons):
\begin{equation}
\label{vn}
v_n(p_T,y,j)\equiv\langle\cos n\phi\rangle =
{\displaystyle\int_0^{2\pi} \cos n\phi {\displaystyle 
dN_j\over\displaystyle d^3{\bf p}}d\phi\over
\displaystyle\int_0^{2\pi}{\displaystyle dN_j\over\displaystyle
 d^3{\bf p}}d\phi}
\end{equation}
where the brackets denote an average value over many events with 
approximately the same impact parameter, 
and $p_T$ and $y$ are the transverse momentum and rapidity 
of the particle. 
Since the system is symmetric with respect to the reaction plane
for spherical nuclei, $\langle\sin n\phi\rangle$ vanishes. 
The goal of the flow analysis is to extract $v_n$ from the data. 
The coefficients 
$v_1$ and $v_2$ are usually called directed and elliptic flow, 
respectively~\cite{Olli98}. 

Since the reaction plane is not known experimentally, one can 
only measure relative azimuthal angles between the outgoing 
particles. 
In particular, one measures 
the Fourier coefficients of the relative azimuthal distribution
between two species of particles $j$ and $k$
\begin{equation}
\label{defcn}
c_n^{\rm measured}(p_{T1},y_1,j;p_{T2},y_2,k)
\equiv \langle\cos n(\phi_1-\phi_2)\rangle
={\displaystyle\int\!\!\!\displaystyle\int{\cos n(\phi_1-\phi_2) 
{\displaystyle dN_{jk}\over \displaystyle d^3{\bf p}_1 
d^3{\bf p}_2}d\phi_1d\phi_2}
\over\displaystyle\int\!\!\!\displaystyle\int{{\displaystyle dN_{jk}
\over \displaystyle 
d^3{\bf p}_1d^3{\bf p}_2}d\phi_1d\phi_2}}.
\end{equation}
The two-particle distribution can generally be expressed as the 
sum of an uncorrelated distribution and two-particle correlations:
\begin{equation}
\label{defc2}
{dN_{jk}\over d^3{\bf p_1}d^3{\bf p_2}}=
{dN_j\over d^3{\bf p_1}}{dN_k\over d^3{\bf p_2}}
\left(1+C_{jk}({\bf p}_1,{\bf p}_2)\right)
\end{equation}
where $C_{jk}({\bf p}_1,{\bf p}_2)$ is the two-particle connected 
correlation function. 
Using this decomposition, one can write $c_n^{\rm measured}$ as the sum 
of two terms:
\begin{equation}
\label{decomposition}
c_n^{\rm measured}(p_{T1},y_1,j;p_{T2},y_2,k)= 
c_n^{\rm flow}(p_{T1},y_1,j;p_{T2},y_2,k)+
c_n^{\rm non-flow}(p_{T1},y_1,j;p_{T2},y_2,k)
\end{equation}
where the first term is due to flow:
\begin{equation}
\label{flo1}
c_n^{\rm flow}(p_{T1},y_1,j;p_{T2},y_2,k)=
v_n(p_{T1},y_1,j)v_n(p_{T2},y_2,k)
\end{equation}
while the remaining term comes from direct two-particle correlations:
\begin{equation}
\label{cnnonflow}
c_n^{\rm non-flow}(p_{T1},y_1,j;p_{T2},y_2,k)=
{\displaystyle\int\!\!\!\displaystyle\int{\cos n(\phi_1-\phi_2) 
C_{jk}({\bf p}_1,{\bf p}_2){\displaystyle dN_j\over \displaystyle d^3{\bf p}_1 }
{\displaystyle dN_k\over \displaystyle d^3{\bf p}_2 }d\phi_1d\phi_2}
\over\displaystyle\int\!\!\!\displaystyle\int{{\displaystyle 
dN_{jk}\over \displaystyle 
d^3{\bf p}_1d^3{\bf p}_2}d\phi_1d\phi_2}}.
\end{equation}

From the flow term (\ref{flo1}), one can calculate the Fourier 
coefficient $v_n$, up to a sign (see \cite{nous} for details):
\begin{equation}
\label{flo3}
v_n (p_{T},y,j)=
\pm{c_n^{\rm flow}(p_{T},y,j;{\cal D})\over 
\sqrt{c_n^{\rm flow}({\cal D};{\cal D})}}
\end{equation}
where ${\cal D}$ denotes the detector used to 
estimate the reaction plane, $c_n(p_{T1},y_1,j;{\cal D})$ 
the average value of $c_n$  over $(p_{T2},y_2,k)$ 
in ${\cal D}$, and $c_n({\cal D};{\cal D})$ 
the average over both $(p_{T1},y_1,j)$  and $(p_{T2},y_2,k)$. 

Experimental analyses usually assume that the only azimuthal 
correlation between outgoing particles is due to their correlation 
with the reaction plane, i.e. they neglect $c_n^{\rm non-flow}$. 
This means that the published flow data are given by 
Eq.(\ref{flo3}) with $c_n^{\rm flow}$ replaced by the 
total correlation $c_n^{\rm measured}$; in turn, 
the total correlation can be reconstructed from the published flow 
data using Eq.(\ref{flo1}). 

We shall see in section \ref{s:sources} that 
$c_n^{\rm non-flow}$ is typically of order $1/N$ 
where $N$ denotes the number of outgoing particles. 
In a central Pb--Pb collision at SPS, $N\sim 2500$, so that one 
expects $c_n^{\rm non-flow}\sim 4\times 10^{-4}$. 
This is certainly a weak effect. 
However, the total azimuthal correlation is also weak: 
since the directed and elliptic flow at SPS measured by NA49 
are of the order of $3\%$ \cite{Data}, one obtains from 
Eq.(\ref{flo1}) 
(since it is assumed that all the measured correlation is from flow)
$c_n^{\rm measured}\sim 9\times 10^{-4}$, 
only a factor of 2 larger than non-flow correlations. 
Therefore one cannot neglect the latter. 
In order to take them into account, 
one must estimate the various contributions to $c_n^{\rm non-flow}$, 
subtract them from $c_n^{\rm measured}$ in 
(\ref{decomposition}) to isolate 
$c_n^{\rm flow}$, and calculate the resulting $v_n$ using (\ref{flo3}).  
In \cite{nous}, we applied this procedure to the NA49 data on pion 
flow \cite{Data}, taking into account correlations due to the HBT effect. 
In section \ref{s:spsdata}, we consider in addition the effects of momentum 
conservation and resonance decays. 

\section{Sources of non-flow correlations}
\label{s:sources}

Several physical effects yield two--particle correlations, which
contribute to the non-flow correlation 
defined in (\ref{cnnonflow}).
In this section, we discuss momentum conservation, resonance 
decays, quantum HBT correlations, and final state interactions, 
and we estimate their respective contributions to 
$c_n^{\rm non-flow}$. 

\subsection{Transverse momentum conservation}

Global momentum conservation gives a back-to-back correlation between 
the produced particles. 
Since we are interested in azimuthal correlations, we consider
only transverse momentum. 
Its contribution to 
$C_{jk}({\bf p}_1,{\bf p}_2)$, which we denote by 
$C_{jk}^{\Sigma p_T}({\bf p}_1,{\bf p}_2)$, is calculated in 
Appendix \ref{s:apA} in the case where there is no other 
correlation between particles. The result is 
\begin{equation}
\label{cmomentum}
C_{jk}^{\Sigma p_T}({\bf p}_1,{\bf p}_2)
= -{2{\bf p}_{T1}\cdot{\bf p}_{T2}\over\langle\sum p_T^2\rangle}
\end{equation}
where the sum in the denominator runs over all the particles 
emitted in the collision. 
This correlation is clearly of order $1/N$. 
The Fourier coefficient $c_n^{\Sigma p_T}$ is 
obtained by  inserting (\ref{cmomentum}) in Eq.(\ref{cnnonflow}). 
Neglecting azimuthal anisotropies of the one-particle distribution,
which are of the order of a few percent~\cite{Data}, 
one finds a result which is non-vanishing only for $n=1$:
\begin{equation}
\label{c1pt}
c_1^{\Sigma p_T}(p_{T1},y_1,j;p_{T2},y_2,k)=
-{p_{T1}p_{T2}\over\langle\sum p_T^2\rangle}.
\end{equation}
Since $c_n^{\Sigma p_T}$ vanishes for $n\not= 1$, the 
effects of momentum conservation must be taken into 
account only in the measurement of directed flow. 
Eq.(\ref{c1pt}) allows to calculate the correlation 
arising from momentum conservation, as soon as the value of 
the denominator $\langle\sum p_T^2\rangle$ is known. 
This quantity is 
evaluated in Appendix \ref{s:apB} in the case of Pb--Pb 
collisions at SPS. The value (\ref{pt2_val}) is probably 
overestimated, which means that we underestimate 
the correction due to momentum conservation in the numerical
calculations of section \ref{s:spsdata}. 
Eq.(\ref{c1pt})  
shows that the azimuthal correlations between two particles 
due to momentum conservation 
are independent of the rapidities $y_1$ and $y_2$ 
and of the particle species $j$ and $k$, but 
increase linearly with transverse momentum. 
The effect is therefore stronger for heavier particles, 
which have larger transverse momenta, as for instance protons. 
The centrality dependence is also easily determined:
since transverse momentum spectra depend weakly on the centrality
\cite{pt,sikler99},
$\langle\sum p_T^2\rangle$ scales like the multiplicity $N$,
so that $c_1^{\Sigma p_T}$ scales like $1/N$. 
Note, finally, that momentum conservation does not contribute 
if the window used for the reaction plane determination is 
symmetric around mid-rapidity~\cite{DANI2}. 

\subsection{Resonance decays}
\label{s:resonances}

We now discuss the azimuthal correlation between the decay 
products of a resonance. It is known that 
many pions originate from $\rho\to\pi\pi$ decays, and that
many nucleons which are excited into $\Delta$ 
resonances\cite{Delta,Bass98}, 
which then decay into a nucleon and a pion. 
When pions are used to determine the reaction plane, 
$\rho$ (resp. $\Delta$) decays must be taken into account 
in the analysis of the pion (resp. proton) flow; when 
protons are used, $\Delta$ decays must be taken into account 
in the analysis of the pion flow. 
Resonances with higher masses are less abundant and will not 
be considered in this paper; however, they may also give 
rise to sizeable correlations. 

The contribution of a resonance decay, say, $\Delta\to p\pi$, to 
$c_n^{\rm non-flow}$ is the product of two factors: 
the first factor is the average value of $\cos n(\phi_1-\phi_2)$
where $\phi_1$ and $\phi_2$ are the azimuthal angles of a
proton and a pion originating from the same $\Delta$; 
the second factor is the probability that a given proton 
and a given pion originate from the same $\Delta$. 
This probability is hard to estimate. 
There are recent measurements of $\rho^0$ 
\cite{rho} and $\Delta^{++}$~\cite{Delta} multiplicities,  
but this is not sufficient: 
indeed, due to the short resonance lifetimes, 
the decay products may interact again after they have been 
produced, in a way which depends on the detailed collision 
dynamics. 
Microscopic models of heavy ion collisions (see \cite{Pang} for a 
review) 
could be used to calculate correlations from resonance decays, but  
the discussion in this paper will remain at a more qualitative 
level.

\begin{figure}[htb]
\begin{center}
\includegraphics[width=0.47\linewidth]{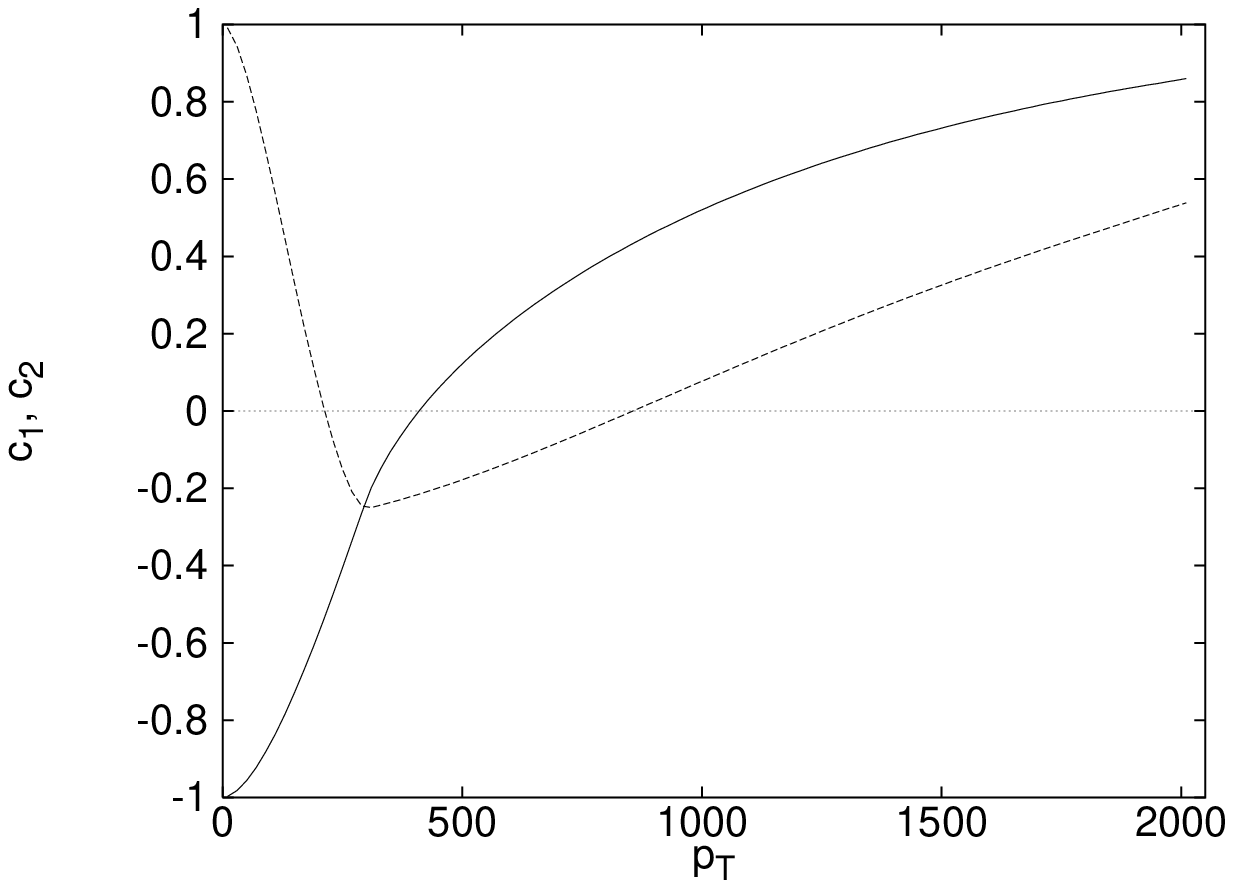}
\hspace{.5cm}
\includegraphics[width=0.47\linewidth]{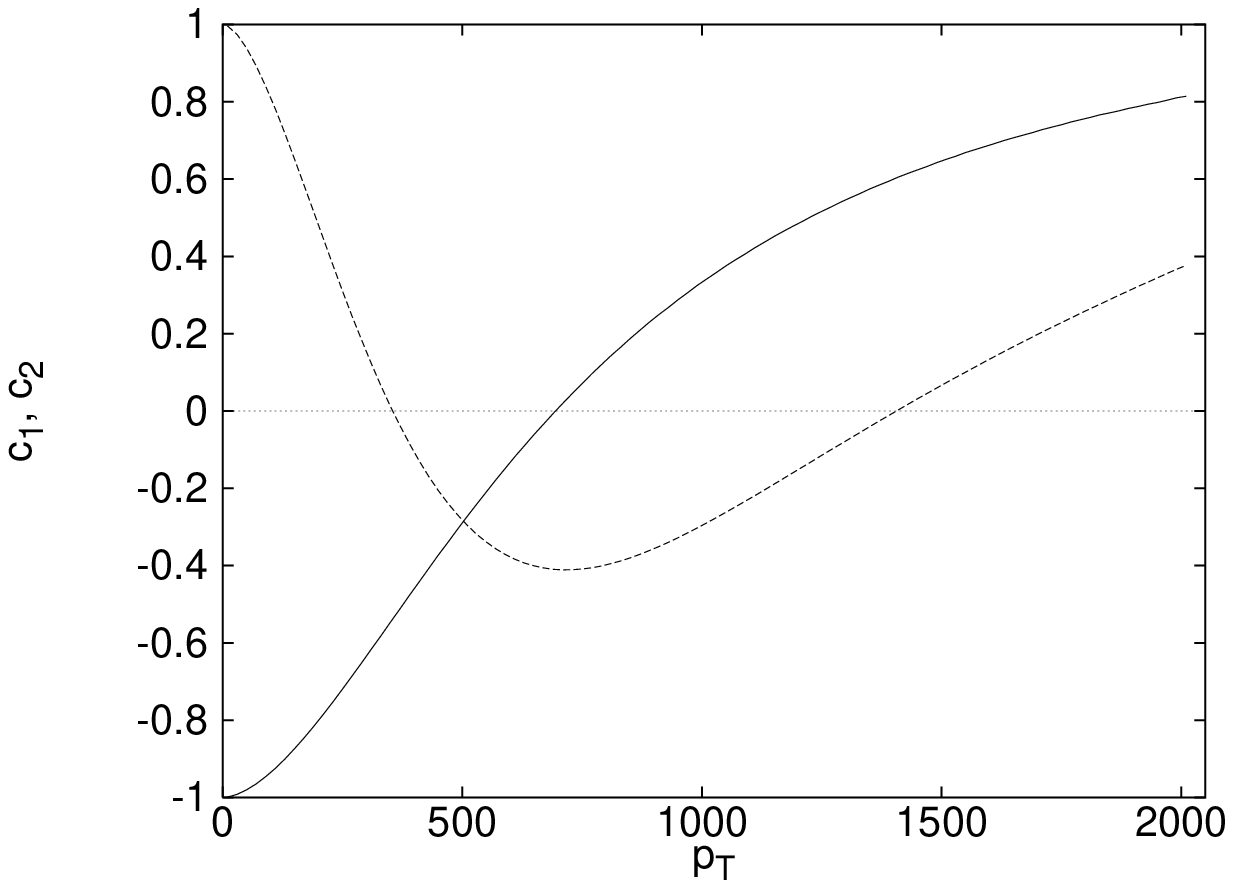}
\end{center}
\caption{
Average value of $\cos(\phi_1-\phi_2)$ (full curve) and 
$\cos 2(\phi_1-\phi_2)$ (dashed curve), where $\phi_1$ and
$\phi_2$ are the azimuthal angles of the decay products of 
$\Delta\to p\pi$ (left) or $\rho\to\pi\pi$ (right), as a function 
of the transverse momentum of the decaying particle in MeV/c. }
\label{fig:deltapt}
\end{figure}

\begin{figure}[htb]
\begin{center}
\includegraphics[width=0.47\linewidth]{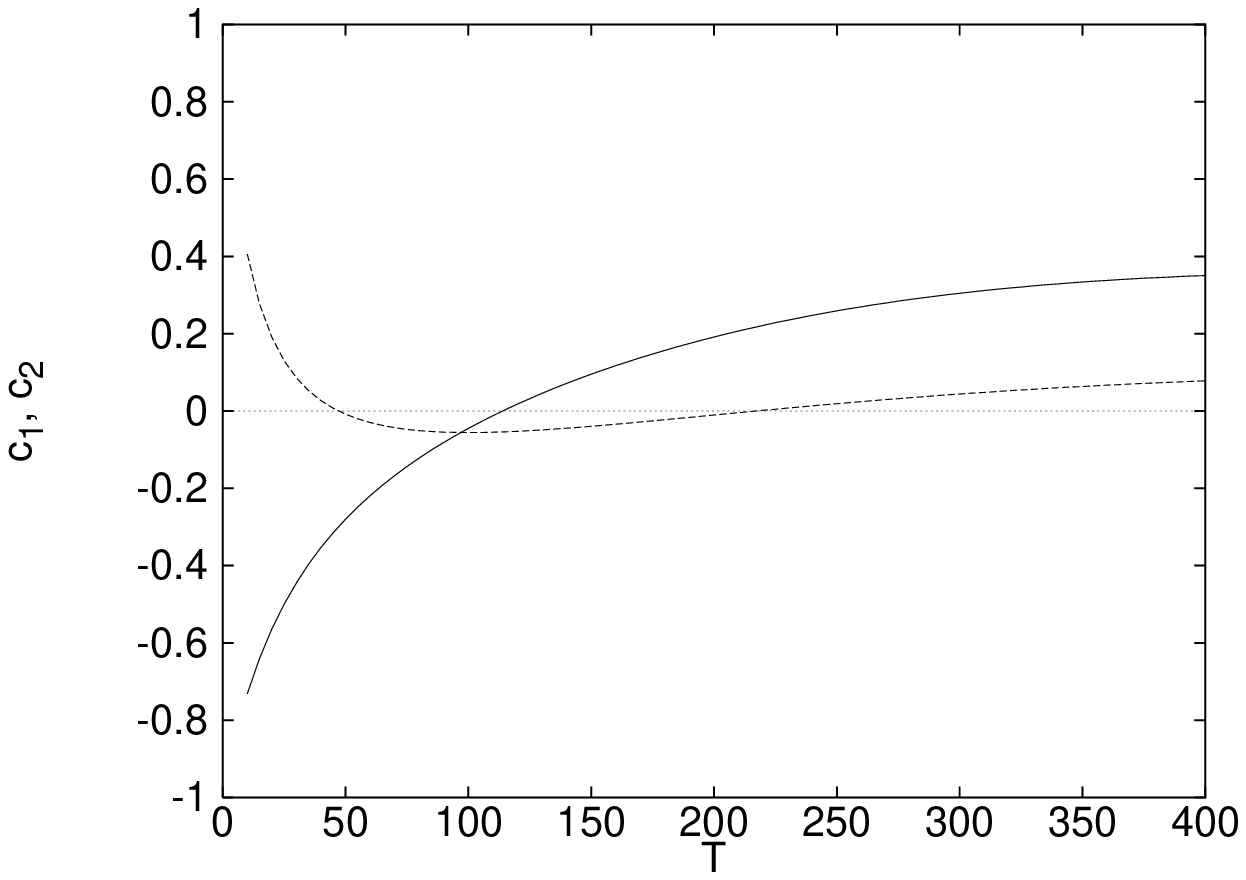}
\hspace{.5cm}
\includegraphics[width=0.47\linewidth]{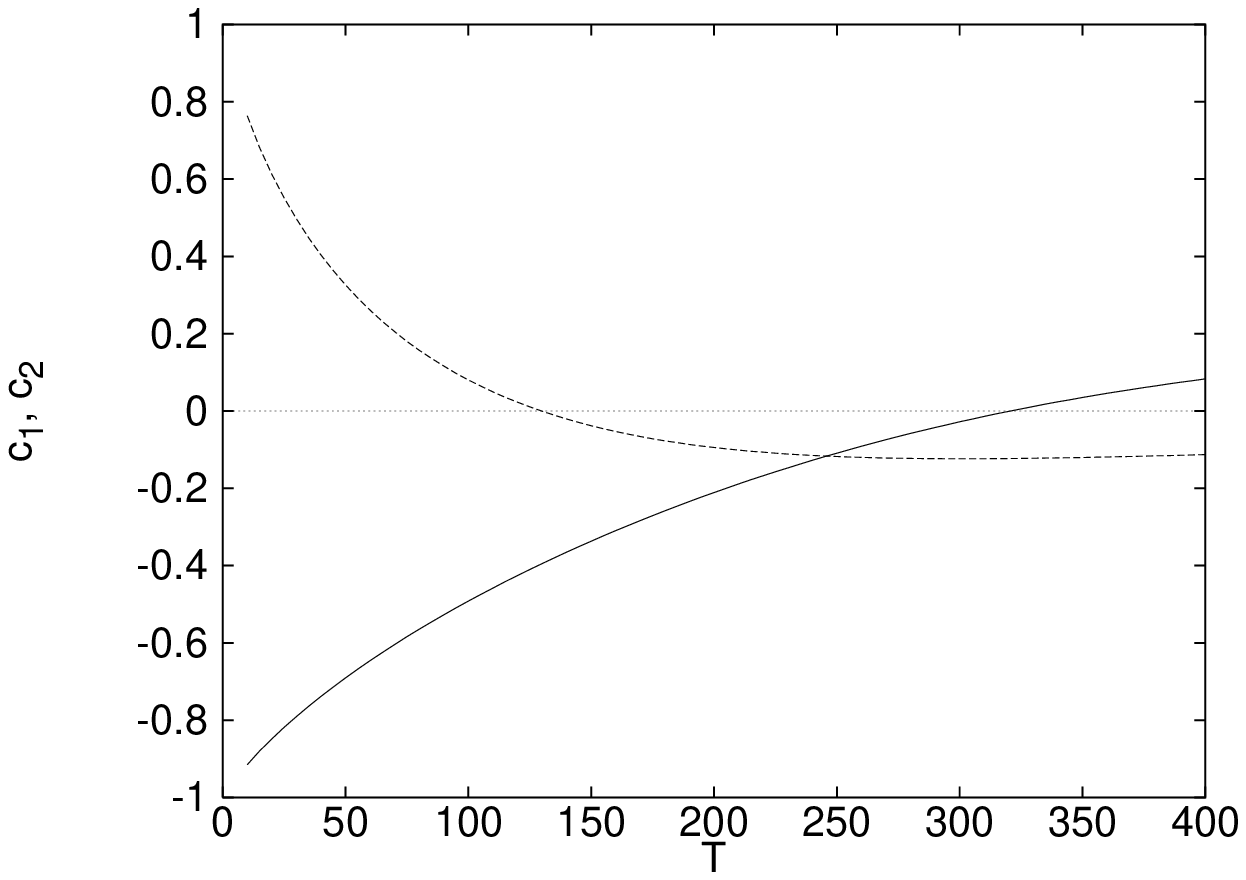}
\end{center}
\caption{
Same as Fig.1, as a function of the inverse slope parameter
 $T$ of the resonance distribution in MeV. Left: $\Delta\to p\pi$; 
Right: $\rho\to\pi\pi$.}
\label{fig:deltaT}
\end{figure}

Generally, in a two-particle decay, 
the azimuthal correlation depends on the relative magnitude of 
the velocity $V$ of the decaying particle and the velocities 
$V_1$ and $V_2$ of the products in the rest frame of the 
decaying particle:
in particular, the azimuthal angles of the decay products 
are equal ($\phi_1=\phi_2$) in the limit 
where $V_1=V_2=0$, while they are back-to-back 
($\phi_1=\phi_2+\pi$) if $V=0$. 
This is illustrated in Fig.\ref{fig:deltapt}, which displays 
the average value of 
$\langle\cos(\phi_1-\phi_2)\rangle$ and 
$\langle\cos 2(\phi_1-\phi_2)\rangle$ for the decays $\Delta\to
p\pi$ and $\rho\to\pi\pi$, 
as functions of the transverse momentum of the decaying resonance. 
Assuming that the transverse momentum distribution is exponential 
in the transverse mass $m_T=\sqrt{p_T^2+m^2}$, 
i.e. $dN/d^2{\bf p}_T\propto \exp(-m_T/T)$ (see
Eq.(\ref{mtspectrum})), one can average the 
correlations displayed in Fig.\ref{fig:deltapt} over $p_T$. 
The result is shown in Fig.\ref{fig:deltaT}, as a function 
of the inverse slope parameter $T$.
Taking for $\Delta$ the same value of $T$ as for protons, 
i.e. $T\simeq 300$~MeV, one sees that the contribution to 
$c_1$ is large and positive, while the contribution to $c_2$ 
is close to zero. In the case of $\rho$ decays, assuming 
that $T$ varies linearly with the mass \cite{Stachel}
one may choose $T\simeq 270$~MeV, which results in 
both $c_1$ and $c_2$ slightly negative. 

In order to obtain the contribution of $\Delta$ decays to 
$c_n^{\rm non-flow}$, one must multiply the correlation displayed in 
Fig.(\ref{fig:deltapt}) by the 
probability $p_\Delta$ that a given pion and a given proton originate 
from the same $\Delta$, as explained above. 
Let $\alpha$ denote the fraction of nucleons are excited into $\Delta$ 
resonances. 
Then, the probability $p_\Delta$ is 
$\alpha/N_\pi$ where $N_\pi$ is the number of pions. From isospin 
symmetry, one expects this probability to be the same for neutral 
and charged pions. Since most particles are pions, $N_\pi\sim N$. 
If $\alpha$ is of order unity, we obtain finally 
$c_n^{\Delta\to p\pi}\sim 1/N$. 
The centrality dependence of this correlation is also 
in $1/N$ if the relative abundance of $\Delta$ is 
independent of centrality. 
Similar arguments hold for the correlations due to $\rho$
decays. 

Note that a way to eliminate correlations from $\Delta$ 
decays experimentally would be to measure protons and pions in 
widely separated rapidity windows. Indeed, the maximum rapidity 
difference between the decay products of a $\Delta$
is $1.5$. In the case of the $\rho\to\pi\pi$ decay, the 
maximum rapidity difference between the outgoing pions 
is $3.3$, so that the corresponding correlation 
is harder to eliminate. 

\subsection{Small angle correlations}

Finally, there are correlations between particles with 
low relative velocities: quantum HBT correlations between 
identical particles, Coulomb interactions between charged 
particles, and strong interactions.

Azimuthal correlations between identical particles 
due to the HBT effect were studied in detail in \cite{nous}. 
The Fourier coefficient $c_n^{\rm HBT}({\bf p_1},{\bf p_2})$ 
is of order unity if the relative momentum $|{\bf p}_1-{\bf p}_2|$ is 
smaller than 
$\hbar/R$, where $R$ is the size of the particle source. 
It depends weakly on $n$. 
The order of magnitude of $c_n^{\rm HBT}$ integrated over phase 
space is therefore $c_n^{\rm HBT}\sim (\hbar/Rp)^3$, where $p$ 
denotes a typical momentum. 
In the case of a high temperature massless pion gas, 
$p$ is of order $T/c$, while the particle density 
$N/R^3$ is of order $(T/\hbar c)^3$, 
so that $c_n^{\rm HBT}$ is again of order $1/N$ \cite{Mrow}. 
One would expect the centrality dependence of $c_n^{\rm HBT}$
to follow the same behavior. 
However, the measured HBT radii vary more slowly than 
$N^{1/3}$~\cite{Rayons1}. In collisions with sulphur projectiles, 
a dependence in $R\propto N^\alpha$ was found with 
$\alpha\simeq 0.2$~\cite{Rayons2}. 
If the same behavior holds for Pb--Pb collisions, 
the dependence of azimuthal HBT correlations with centrality 
will be in $N^{-0.6}$. 
Finally, $c_n^{\rm HBT}({\cal D})$ is proportional
to the one--particle momentum distribution, hence it is 
larger at low $p_T$. 

More generally, all particles with low relative velocities 
undergo final state interactions, strong and/or electrostatic
\cite{FSI}. 
One defines the invariant momentum as 
$Q_{\rm inv}=2\mu v_{\rm rel}$, where $\mu=m_1m_2/(m_1+m_2)$ is 
the reduced mass and $v_{\rm rel}=\sqrt{1-m_1^2m_2^2/(P_1\cdot
P_2)^2}$ is the relative velocity of the two particles 
with 4-momenta $P_1$ and $P_2$. Then 
the correlations due to strong interactions are sizeable only
in a limited $Q_{\rm inv}$ range, independent of the size of 
the interaction region. 
For instance, proton--proton correlations, which are dominated 
by strong interactions, are maximum for 
$Q_{\rm inv} \simeq 20$~MeV/c, and negligible for $Q_{\rm inv} 
\gtrsim 40$~MeV/c, for any projectile and target\cite{Bauer}.
On the other hand,
the range over which Coulomb correlations are significant 
depends on the system size. For Pb--Pb collisions at SPS, 
$p-\pi^+$ correlations, 
which are dominated by the Coulomb repulsion, extend up to 
$Q_{\rm inv}\simeq 50$~MeV/c~\cite{Lednicky}.
Both Coulomb and strong interactions
give rise to small angle azimuthal correlations, similar 
to those due to the HBT effect. 
Thus, the resulting contribution to $c_n^{\rm non-flow}$, 
denoted by $c_n^{\rm FSI}$, 
is expected to be weakly dependent on the order $n$ of the Fourier 
component, and $c_n^{\rm FSI}(p_T,y,{\cal D})$ is 
roughly proportional to the one--particle distribution.

The centrality dependence differs for Coulomb and strong interactions. 
For strong interactions, the strength of the correlation at a given 
$Q_{\rm inv}$ varies like $1/V$, where 
$V$ is the volume of the system~\cite{FSI}. This is due to the fact 
that the range of the interaction is much smaller than the size of the 
system. 
Thus the centrality dependence follows a $1/N$ 
behavior. 
For Coulomb interactions, which are long--ranged, the $N$ 
dependence is weaker. We do not attempt to evaluate this dependence. 
By analogy with the Coulomb potential, it might 
vary like $1/L$, with $L$ the size of the interaction region,
i.e. like $1/N^{1/3}$. 
 
Both HBT and final state interactions can be eliminated by 
studying the correlation between particles widely separated in 
phase space, so that their invariant relative momentum is much larger 
than, typically, 50~MeV/c. 

\subsection{Summary}

The properties of the three main 
types of two--particle correlations studied above are 
summarized in Table I. 
Note that while the average over all phase space of the 
three is of order $1/N$, 
they can be much larger in definite regions of phase space, 
for example at low relative momentum for HBT correlations. 
Note also that this catalogue of two-particle correlations 
may not be exhaustive. 

As we have seen, two--particle correlations are generally of order 
$1/N$,  much smaller than unity. 
We therefore assume that the contributions of the various effects can 
simply be added. 
If one studies the correlation between identical pions, for instance, 
there are contributions to $c_n^{\rm non-flow}$ 
from momentum conservation, denoted
by $c_n^{\Sigma p_T}$, from $\rho\to\pi\pi$ decays, denoted 
by $c_n^\rho$, from HBT correlations, denoted by $c_n^{\rm HBT}$, 
and from final state interactions, denoted by 
$c_n^{\rm FSI}$. The resulting non-flow correlation is 
\begin{equation}
\label{hbtpt}
c_n^{\rm non-flow}=
c_n^{\Sigma p_T}+c_n^\rho+c_n^{\rm HBT}+c_n^{\rm FSI}.
\end{equation}

\begin{tabular}{|l|c|c|c|}
\hline
& Momentum conservation & Resonance decays
 & Small angle correlations
 \\ \hline
Which pairs? & 
any & $|y_1 - y_2| < 1.5$
 ($\Delta$)
& low relative velocities \\
&& $|y_1 - y_2| < 3.3$ ($\rho$) & identical particles (HBT) 
\\ \hline
$n$ dependence & 
$n=1$ only & non-trivial & weak 
\\ \hline
centrality dependence & 
$1/N$ & $1/N$
& $1/N^{0.6}$~? (HBT)\\
&&& $1/N$ (strong)\\
&&& $1/N^{1/3}$~? (Coulomb)
 \\ \hline
$p_T$ range & 
high $p_T$ & non-trivial & low $p_T$ 
 \\ \hline
under control? & 
calculated & simulations, 
microscopic models & measured 
\\ 
\hline
\end{tabular}

\medskip
\centerline{TABLE I: Properties of the various sources of two--particle 
azimuthal correlations}

\section{Application to SPS data}
\label{s:spsdata}

The most detailed flow analyses in Pb--Pb collisions at SPS were 
performed by the NA49 and WA98 collaborations. 
NA49 estimates the reaction plane using pions in the forward 
hemisphere, while 
WA98 uses protons and fragments close to the target rapidity. 
In this section, 
we discuss to what extent non--flow correlations may contribute 
to the measured azimuthal correlations, which are assumed to be 
solely due to flow in both experiments. 
In the case of the NA49 data, we shall explicitly subtract 
the various non-flow correlations, following the method outlined
in section \ref{s:sec2}, and see what the ``true'' flow might be. 
Our discussion of WA98 data will remain at a semi-quantitative 
level due to the complex acceptance of the Plastic Ball detector 
used for the flow analysis. 

\subsection{NA49 Data}

The NA49 experiment at CERN measures the directed and elliptic flow of pions 
and protons in Pb--Pb collisions at 158~GeV per nucleon~\cite{Data}. 
Charged pions are used to estimate the reaction plane, 
in the kinematic window $4<y<6$ and $0<p_T<600$~MeV/c for directed 
flow,  $3.5<y<5$ and $0<p_T<2000$~MeV/c for elliptic flow \cite{private}. 
The azimuthal distribution of identified particles (protons and 
charged pions) is then measured with respect to this reaction plane. 
We now show the effect of taking into account the various sources 
of non--flow correlations discussed in the previous section. 

\begin{figure}[htb]
\begin{center}
\includegraphics{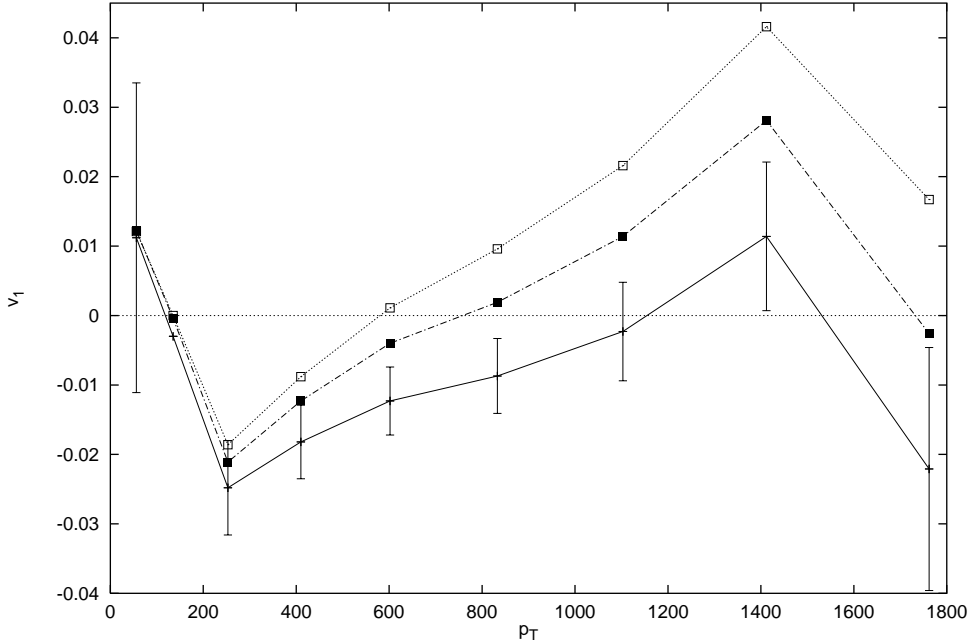}
\end{center}
\caption{Directed flow $v_1$ of protons, integrated over the range 
$3<y<6$, as a function of the transverse momentum $p_T$, measured by 
NA49 (open squares), taking into account momentum conservation 
only (with error bars), and with a simulation of correlations 
from $\Delta$ decays (filled squares).
Error bars are taken from experiment.}
\label{fig:protons}
\end{figure}

We begin with the directed flow of protons. 
It is obtained from from Eq.(\ref{flo3}), where ${\cal D}$ 
refers to charged pions in the above mentioned kinematic window. 
The numerator in this equation corresponds to the correlations between 
protons and the pions in ${\cal D}$, while the denominator 
corresponds to the correlations between the pions in ${\cal D}$.
The arbitrary sign in Eq.(\ref{flo3}) is chosen 
negative by NA49, in order to obtain a positive $v_1$ for protons 
(see Fig.\ref{fig:protons}), 
as usually assumed at high energies in the forward rapidity 
region\cite{DANI}.

The correction to the denominator in Eq.(\ref{flo3}) from non-flow
correlations will be discussed below, when we discuss the pion flow. 
It is a small correction, of a few percent.  
The numerator of Eq.(\ref{flo3}) gets contributions from 
momentum conservation, $\Delta\to p\pi$ decays, and final state 
interactions. 

We first discuss momentum conservation. The corresponding
azimuthal correlation is given by Eqs.(\ref{c1pt}) and 
(\ref{pt2_val}). 
Since it is negative, once  it is subtracted from the measured 
correlation (\ref{decomposition}), this gives a positive correction 
to the correlations due to flow $c_1^{\rm flow}$. 
Because of the negative sign in (\ref{flo3}), the correction to $v_1$ 
is finally negative. 
As can be seen in Fig.\ref{fig:protons}, 
the correction is so large that the sign of the proton directed flow 
is now negative for almost all $p_T$. 
Now, the sign of the proton directed flow is particularly important 
since it is used to fix the arbitrary sign in Eq.(\ref{flo3}), 
which is chosen so that $v_1>0$ in the forward rapidity region for protons. 
The surprising result is that the directed flows of pions and protons 
now have the same sign. This is contrary to the theoretical
expectation that nucleons and pions flow in opposite directions, 
because of pion rescattering
\cite{Bass,Li}, and to other measurements at AGS
\cite{E877} and at SPS \cite{WA98}. 

Since the correlation between pions and protons from 
$\Delta\to p\pi$ decays is positive 
(see Fig.\ref{fig:deltaT}), it may restore the original 
sign of the proton directed flow $v_1$. 
As explained in the previous section, an accurate computation 
of this correlation would require a microscopic model. 
We present in Fig.\ref{fig:protons} a crude estimate of the 
effect, based on a simple Monte-Carlo calculation taking into 
account the acceptance windows of NA49 for protons and pions. 
We do not take into the possible rescatterings of the decay 
products. 
Our assumptions are the following:
50\% of the outgoing protons originate from $\Delta$ decays, 
as suggested by simulations based on the UrQMD model \cite{Bass98}. 
We take into account the fact that only $2/3$ of $\Delta$ decays 
yield a charged pion. 
The mass distribution of $\Delta$'s follows a lorentzian, 
their rapidity distribution is flat between 0 and 5.8 
(i.e. between the target and projectile rapidities), and their 
transverse momentum distribution is exponential in $m_T$ 
(see Eq.(\ref{mtspectrum}))
with an inverse slope parameter $T=300$~MeV, as for protons. 
One sees in Fig.\ref{fig:protons} that correlations from $\Delta$ decays
can indeed restore the positive sign for the proton directed flow, since 
their contribution is almost as large as that from momentum 
conservation in absolute value.

In principle, one should also take into account final state
interactions. 
Since the pions used to determine the reaction 
plane are approximately half $\pi^+$ and half $\pi^-$, Coulomb 
effects cancel. 
Correlations between pions and protons due to strong interactions 
are small and should not contribute significantly to 
$c_n^{\rm measured}$. 

\begin{figure}[htb]
\begin{center}
\includegraphics{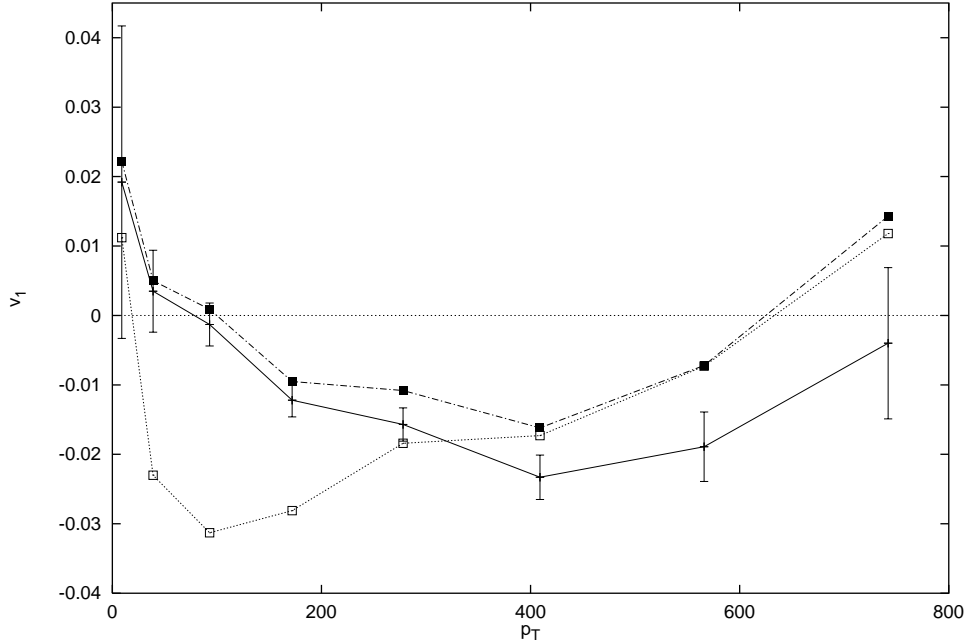}
\end{center}
\caption{Directed flow $v_1$ of pions, integrated between $4<y<5$, as 
a function of the transverse momentum $p_T$ in MeV/c, 
measured by NA49 (open squares), after 
subtraction of HBT correlations (filled squares), and after subtraction 
of both HBT and transverse momentum conservation correlations (with 
error bars).}
\label{fig:fig1}
\end{figure}

We now turn to the directed flow of charged pions.
Since pions are also used to estimate the reaction plane, 
pion--pion correlations are involved in both the numerator 
and the denominator of Eq.(\ref{flo3}). 
Following Eq.(\ref{hbtpt}), we must take into account 
momentum conservation, $\rho$ decays, the HBT effect,  
and final state interactions. 
We neglect final state interactions for the same reasons 
as for protons. 
According to Fig.\ref{fig:deltaT}, the correlation 
from $\rho$ decay is small. We also neglect it. 
Three sets of points are displayed in Fig.\ref{fig:fig1}:
NA49 data without correction, with correction for HBT correlations 
(taken from \cite{nous}) and with correction for HBT 
and momentum conservation. 
While the correction due to the HBT effect is important at low $p_T$, 
momentum conservation changes the pion flow at high $p_T$;
the latter effect is less important than for protons, 
because pions have lower transverse momenta. 
However, the positive pion flow observed by NA49 
at high $p_T$ seems to be explained by momentum 
conservation. 

Let us say a few words of elliptic flow. 
Here, momentum conservation does not contribute, as discussed
in the previous section. 
For protons, $\Delta$ decays probably give a negligible
contribution to $c_2^{\rm measured}$, as explained in section 
\ref{s:resonances}. For pions, HBT correlations significantly 
reduce elliptic flow at low $p_T$, as shown in \cite{nous}. 
Here, $\rho$ decays should be taken into account. 
As discussed in section \ref{s:resonances}, they give a negative 
correlation, 
i.e. opposite to the correlation due to flow. Therefore, 
taking $\rho$ decays into account would lead to a (slight) 
increase of the pion $v_2$. 

Finally, the centrality dependence of the directed and elliptic
flow of pions has been studied recently \cite{posk99}. 
It is striking to note that both increase continuously as 
the reaction becomes more peripheral, which is the behavior 
expected for non--flow correlations. 
We have shown that at least a fraction of the measured 
correlation, which is interpreted as flow, is due to 
non--flow correlations. 
Even if it is only a small fraction, it will increase with 
impact parameter and eventually dominate for the most 
peripheral collisions. It is thus likely that the measurements 
of flow at large impact parameter are contaminated by 
non--flow phenomena. 
However, none of the above mechanisms can explain the observed $y$ 
dependence of directed flow, which increases (in absolute value) 
up to the projectile rapidity for peripheral collisions. 

\subsection{WA98 Data}

We shall now discuss the results obtained by 
the WA98 experiment at CERN for the directed flow 
of protons and positive pions \cite{WA98}. 
There, the reaction plane is estimated using protons 
and fragments detected by the Plastic Ball, in the target 
rapidity region $-0.6<y<0.3$. 
One then measures the azimuthal angle of a positive pion
or a proton detected in the Plastic Ball 
with respect to this estimated plane. 
If the particle is a proton, it is not used in the determination 
of the reaction plane to avoid autocorrelations. 
Two points closer to midrapidity 
are obtained using the tracking system, for which the flow 
is zero within error bars; we shall not consider these two points 
here. Contrary to the NA49 analysis where pions were used to 
estimate the reaction plane, 
the flow of protons (resp. pions) is now determined from 
proton--proton (resp. pion--proton) azimuthal correlations. 
The value of $v_1$ is still obtained from Eq.(\ref{flo3}), 
where the arbitary sign is again chosen negative, 
so that the directed flow of protons be negative, 
as it should in the backward rapidity region. 
On the other hand, the directed flow of pions is positive, 
i.e. pions flow opposite to the protons. 

WA98 measures the directed flow $v_1$ in the target fragmentation
region. 
Since $v_1$ vanishes at midrapidity, it 
is expected to be larger there than in the more central rapidity 
region, where NA49 is working. 
Indeed, the measured values of $v_1$ are larger for WA98 than for NA49. 
For semi-central collisions, $|v_1|$ is about $0.1$ for 
both pions and protons.
Following Eq.(\ref{flo1}), this  corresponds to an azimuthal 
correlation $|c_1^{\rm measured}|\sim 10^{-2}$, either 
between pions and protons or between protons,  
an order of magnitude larger than for NA49. 

The values of $v_1$ are measured as a function 
of centrality. 
For protons, $-v_1$ reaches a maximum of $0.12$ for an impact 
parameter $b\sim 8$~fm, while for pions $v_1$ increases with 
$b$ up to very peripheral collisions, where it reaches values 
as high as $0.3$.
The latter centrality dependence is reminiscent of the $1/N$ 
behavior of non--flow correlations, which suggests that 
part of the measured flow is spurious. 
We thus review the various non--flow correlations listed in 
section \ref{s:sources} to evaluate their contributions. 
We shall see that final state interactions are the only 
significant ones, and that they may explain the pion flow 
at large impact parameter. 

Momentum conservation is less important than for the NA49
experiment. 
The reason is that the particles detected in the Plastic Ball 
have low transverse momenta~\cite{Hubertus}. 
For protons, $p_T$ is typically 400~MeV, resulting in 
$c_1^{\Sigma p_T}$ of the order of $-4\times 10^{-4}$ 
for the proton--proton correlation, following Eq.(\ref{c1pt}). 
This is much smaller than $c_1^{\rm measured}\sim 10^{-2}$. 
For pions, transverse momenta are even smaller. In both 
cases, the contribution of transverse momentum conservation 
to $c_1^{\rm non-flow}$ is negligible. 

Resonance decays will contribute to the pion flow. 
The sign of the correlation due to $\Delta\to p\pi$ depends 
on the inverse slope parameter $T$, as can be seen in
Fig.\ref{fig:deltaT}. 
This quantity is smaller in the fragmentation region than 
in the central rapidity region, 
but probably still larger than 100~MeV. Then, the corresponding 
pion--proton correlation should be positive. It is thus 
opposite to the correlation measured by WA98, and cannot 
explain the observed pion flow. 

Finally, let us discuss the small angle correlations 
due to the HBT effect and final state interactions. 
We shall first summarize the results obtained by the NA49 
experiment for two-particle small angle correlations in the 
central rapidity region. Then we extrapolate these results to 
the fragmentation region covered by the WA98 Plastic Ball in 
order to discuss the correlations measured by WA98.

The NA49 collaboration has measured both proton-proton 
\cite{pp} and proton-$\pi^+$ \cite{Lednicky} small angle 
correlations for central collisions. 
Two-proton correlations, resulting from the interplay of the 
HBT effect and strong and Coulomb interactions, were found to 
be positive, which means that they are dominated by the 
attractive strong interaction. They peak at a relative momentum 
between particles of 20~MeV/c, where they reach $25\%$, and are 
no longer significant as soon as $|{\bf p}_1 - {\bf p}_2| > 
40$~MeV/c. Small angle correlations between positive pions and 
protons are negative, and typically of the order of $-10\%$ for 
invariant momenta smaller than 40~MeV/c, i.e. 
$|{\bf p}_{T\pi}-(m_\pi/m_p){\bf p}_{Tp}|<40$~MeV/c and 
$y_\pi-y_p<40/m_\pi\simeq 0.3$. 

Extrapolating these results to the fragmentation region, 
we can estimate the azimuthal correlation $c_1^{\rm FSI}$
for WA98. 
In the case of proton--proton correlations, 
we need to evaluate the fraction of proton pairs 
having relative momenta less than 40~MeV/c, i.e. with 
$|{\bf p}_{T1}-{\bf p}_{T2}|<\Delta p=40$~MeV/c and 
$y_1-y_2<\Delta p/m_T\simeq 0.04$, with $m_T\simeq 1000$~MeV.  
Since the acceptance window is typically~\cite{Hubertus} 
$p_T^{\min}=200<p_T<p_T^{\max}=600$~MeV/c 
and $y^{\min}=-0.4<y<y^{\max}=0.1$, this fraction 
\begin{equation}
\label{fraction}
{{4\pi\over 3}\Delta p^3\over \pi(p_T^{\max 2}-p_T^{\min 2})
m_T(y^{\max}-y^{\min})}\sim 5\times 10^{-4}.
\end{equation}
With a correlation strength of 20\%, 
this yields a value of $c_1^{\rm FSI}$ of the order of $10^{-4}$, 
much smaller than the measured correlation $c_1^{\rm measured}$. 
Thus, the measurement of the proton flow is not influenced 
by small angle correlations. 

In the case of pion--proton correlations, we first note 
that the velocity windows of pions and protons of the WA98 
Plastic Ball overlap~\cite{Hubertus}, so that small angle 
correlations between pions and protons should be taken into account. 
In fact, the rapidity range is almost the same for pions and
protons, while the transverse momentum range is roughly scaled by 
a factor $m_\pi/m_p$, so that the acceptance windows 
roughly coincide in velocity space. 
Then, for a given pion, the probability that a proton seen in the 
Plastic Ball has an invariant relative momentum 
$|{\bf p}_{T\pi}-(m_\pi/m_p){\bf p}_{Tp}|<\Delta p=40$~MeV/c 
is given by (\ref{fraction}) with $\Delta p$ replaced by 
$(m_p/m_\pi)\Delta p$.  
This gives a factor $(m_p/m_\pi)^3\simeq 200$, compared to 
Eq.(\ref{fraction}), so that the fraction is of the order of 
$10^{-1}$. 
Using the correlation strength measured by NA49, this gives a 
correlation $c_1^{\rm FSI}({\cal D})$ of order 
$-10^{-2}$, comparable to the measured correlation. 
On the one hand, this value is overestimated since the acceptance 
windows for pions and protons do not exactly coincide in velocity 
space.
On the other hand, it is underestimated because the correlations 
are measured by NA49 for 
central collisions, and they are even stronger
for peripheral collisions. 
Therefore it is likely that a large fraction of the 
$\pi^+$ directed flow seen by WA98 is due to the repulsive 
Coulomb interaction between positive pions and protons. 

\section{Discussion}
\label{s:sec5}

We have shown that the assumption that all azimuthal correlations 
between particles are due to flow is no longer valid at SPS 
energies, where other sources of correlations become of 
the same order of magnitude due to the smallness of the 
flow. 
All the effects studied above, momentum 
conservation, resonance decays, HBT correlations, final state interactions,
turn out to give important correlations. 
In particular, we have shown that the effect of momentum 
conservation is so large that it changes the sign of the
proton directed flow measured by NA49: this sign is particularly
important since it is used to define the orientation of the 
reaction plane.

The methods currently used to analyse the flow can be modified 
in order to take into account these additional correlations and 
to subtract them from the measured correlations, 
as explained in section \ref{s:sec2}. 
Correlations from momentum conservation can be calculated. 
They are governed by only one unknown quantity, the sum over all particles 
of squared transverse momenta $\langle \sum p_T^2\rangle$. 
Correlations from the decays of short-lived resonances 
cannot be evaluated so easily. In this paper, we have only given 
semi-quantitative estimates. 
More accurate calculations would require a microscopic 
model of the collision. 
Finally, correlations which affect particles with low relative 
velocities, i.e. those due to the HBT effect and to final state 
(Coulomb, strong) interactions can either be measured independently, 
or eliminated by measuring correlations between particles in widely 
separated rapidity and/or transverse velocity windows. 

However, our list of correlations may not be exhaustive, and one 
cannot exclude that other sources exist which are of the same order 
of magnitude as those studied here. 
There is no systematic way to 
separate the flow and the non--flow contribution to the 
measured azimuthal correlation (\ref{decomposition}). 
One could try to use the factorization property of Eq.(\ref{flo1}) 
to isolate the correlation due to flow. 
However, non-flow correlations may also factorize, as is 
the case for the correlation due to momentum conservation 
(\ref{c1pt}). 

A way to circumvent these difficulties would be to use the remarkable
property of two--particle correlations, that they vary with the 
multiplicity $N$ like $1/N$ (with the possible exception of Coulomb 
interactions and, to a lesser extent, HBT correlations). 
This fact has been noted in section 
\ref{s:sources} for the various correlations studied, but holds 
more generally for any extensive system, and would probably still 
be true for other types of correlations than those listed in 
section \ref{s:sources}. 
This $1/N$ behavior determines the centrality dependence of 
non--flow two--particle correlations. Since the flow is known to vanish for 
central collisions by symmetry, one could measure (and then 
subtract) non-flow correlations by measuring them for central
collisions and then assuming that they vary like $1/N$. 
Unfortunately, it is impossible to select ``true'' central 
collisions (with vanishing impact parameter) experimentally, 
so that this procedure may not be accurate.

Future heavy ion experiments at RHIC and LHC are likely to 
be affected in the same way by non--flow correlations. 
At these energies, directed flow 
is expected to be weaker than at SPS, maybe undetectable. 
On the other hand, elliptic flow should be at least of the same 
magnitude as at SPS. 
Pions should be used to determine the reaction plane since 
they are by far the most abundant. 
As the multiplicity will be higher than at SPS, non-flow correlations 
will be smaller, since they are proportional to $1/N$. 
However, since the detectors will cover only a limited interval in 
rapidity, effects of $\rho$ decays, HBT correlations and final state 
interactions may be important 
and should be taken into account in the measurement of 
the pion elliptic flow. Hard processes could give an additional 
contribution to non--flow correlations: there is a strong azimuthal 
correlation between jet fragments~\cite{leonidov}. 
Clearly, more work is needed, both theoretical and experimental, 
in order to obtain accurate measurements of flow observables at 
ultrarelativistic energies.
Alternative methods, based on multiparticle correlations, are 
currently under study~\cite{nearfuture}.

\bigskip\bigskip
\acknowledgements

We thank the Nuclear Theory group of Brookhaven National Laboratory 
for its hospitality during the period when this work was done.  
We thank Terry Awes, Art Poskanzer, Hubertus Schlagheck and Sergei
Voloshin for detailed explanations concerning the NA49 and WA98 
flow analyses, and Stefan Bass for discussions.

\appendix
\section{Correlations from momentum conservation}
\label{s:apA}

We denote by ${\bf p}_1\cdots {\bf p}_N$ 
the momenta of the $N$ particles emitted in a heavy ion collision 
and by ${\bf p_T}_1\cdots {\bf p_T}_N$ their transverse components. 
Since we are interested in azimuthal correlations, we consider 
only transverse momentum conservation:
${\bf p_T}_1+\dots +{\bf p_T}_N={\bf 0}$. 
We assume for simplicity that this is the only correlation between 
the momenta of the outgoing particles. 
We want to calculate the distribution of ${\bf p}_1,\dots,{\bf p}_k$, 
with $k< N$, defined as:
\begin{equation}
\label{fc}
f_c({\bf p}_1,\dots,{\bf p}_k)\equiv {
\left(\displaystyle\prod_{i=1}^k f({\bf p}_i)\right)\displaystyle\int{
\delta^2({\bf p_T}_1+\dots+{\bf p_T}_N)\prod_{i=k+1}^{N}
\left(f({\bf p}_i)d^3{\bf p}_i\right)}\over \displaystyle
\int{\delta^2({\bf p_T}_1+\dots+{\bf p_T}_N)\prod_{i=1}^{N}
\left(f({\bf p}_i)d^3{\bf p}_i\right)}}
\end{equation}
where $f({\bf p})$ denotes the single particle normalized transverse
momentum distribution. The average transverse momentum is naturally 
assumed to be zero:
\begin{equation}
\langle{\bf p_T}\rangle\equiv\int {\bf p_T} f({\bf p}) d^3{\bf p}={\bf 0}.
\end{equation}

In order to calculate the distribution $f_c$ defined in Eq.(\ref{fc}), 
we make use of the following lemma:
the sum of $M$ uncorrelated momenta 
${\bf P_T}\equiv\sum_{i=1}^M{\bf p_T}_i$ has a gaussian distribution 
if $M$ is large, according to the central limit theorem:
\begin{eqnarray}
\label{pm}
F_M({\bf P_T})&\equiv&\int{\delta^2\left(-{\bf P_T}+
\sum_{i=1}^M{\bf p_T}_i\right)
\prod_{i=1}^M\left( f({\bf p}_i)d^3{\bf p}_i\right)}\cr
&=& {1\over\pi\sigma^2}\exp\left(-{{\bf P_T}^2\over\sigma^2}\right).
\end{eqnarray}
In writing this equation, we have assumed that the transverse momentum 
distribution is isotropic in the transverse plane, i.e. we have
neglected the azimuthal asymmetries (\ref{vn}), which are of the order 
of a few percent. 
The width of the gaussian is 
\begin{equation}
\label{sigma2}
\sigma^2=\langle {\bf P_T}^2\rangle=M\langle p_T^2\rangle. 
\end{equation}

Eqs.(\ref{pm}) and (\ref{sigma2}) allow to write (\ref{fc}) as  
\begin{eqnarray}
\label{fcorrelee}
f_c({\bf p}_1,\dots,{\bf p}_k)&=&
\left(\prod_{i=1}^k f({\bf 
p}_i)\right){F_{N-k}\left(-{\displaystyle\sum_{i=1}^k}
{\bf p_T}_i\right)\over F_N(0)}
\cr
&=&
\left(\prod_{i=1}^k f({\bf p}_i)\right)
{N\over N-k}
\exp\left(-{({\displaystyle\sum_{i=1}^k}
{\bf p_T}_i)^2\over (N-k)\langle {\bf p_T}^2\rangle}\right).
\end{eqnarray}

For $k=1$, Eq.(\ref{fcorrelee}) gives the corrected 
one-particle distribution
\begin{equation}
\label{f1}
f_c({\bf p})=f({\bf p}) \left(1+{1\over N}-{{\bf p_T}^2\over
N\langle p_T^2\rangle}\right)
\end{equation}
where we have expanded the correction to leading order in $1/N$. 
Similarly, one obtains for k=2
\begin{equation}
\label{f2}
f_c({\bf p}_1,{\bf p}_2)=
f({\bf p}_1)f({\bf p}_2)\left(1+{2\over N}-
{({\bf p_T}_1+{\bf p_T}_2)^2\over N\langle p_T^2\rangle}\right).
\end{equation}
The contribution of momentum conservation to the 
correlation function (\ref{defc2}) is thus 
\begin{equation}
C^{\Sigma p_T}({\bf p}_1,{\bf p}_2)={f_c({\bf p}_1,{\bf p}_2)\over 
f_c({\bf p}_1)f_c({\bf p}_2)}-1 
= -{2{\bf p_T}_1\cdot{\bf p_T}_2\over N\langle p_T^2\rangle}
\end{equation}
Similar results were obtained in \cite{DANI2}, Appendix B. 


\section{Estimate of $\sum \langle p_T^2 \rangle$ at SPS}
\label{s:apB}

In a central Pb-Pb collision at SPS, about 680 negatively charged 
particles are emitted \cite{SpectraNA49}. 
If we assume that these are mostly $\pi^-$ and $K^-$ with a ratio 
$\langle K^- \rangle/\langle \pi \rangle \sim 0.09$, where 
$\langle\pi\rangle\equiv(\langle \pi^+ \rangle+\langle \pi^- \rangle)/2$ 
\cite{sikler99}, there are some 625 $\pi^-$ and 55 $K^-$. 
Taking into account the various kaon-to-$\pi$ ratios, as well as the 
expected isospin symmetry between pion species, this gives a total 
of 280 kaons and 1875 pions. With these particles come the 416 
nucleons (we assume there is no fragment), giving about 2500 
particles in all. We shall neglect the contributions to $\sum 
\langle p_T^2 \rangle$  from other particles as e.g. antiprotons.

In order to calculate the average value of the squared transverse 
momentum $\langle p_T^2 \rangle$ for a particular particle species, 
one needs its distribution in transverse momentum and rapidity. 
The $p_T$- and $y$-dependences may factorize, leaving a normalized 
$p_T$ distribution which is parametrized by
\begin{equation}
\label{mtspectrum}
{dN\over d^2{\bf p_T}}= {e^{m/T}\over 2\pi T (m+T)}\exp\left(-{m_T\over
T}\right).
\end{equation}
where $m_T=\sqrt{m^2+p_T^2}$ is the transverse mass, while the inverse 
slope parameter $T$ depends on the particle species and the energy of the 
collision. 
This parametrization can be used for the pions, nucleons and kaons at 
SPS, with $T_\pi=180$ MeV, $T_N=300$ MeV and $T_K=215$ MeV respectively 
\cite{SpectraNA49}. 
These numerical values fit transverse momentum spectra in the 
central rapidity region. Away from mid-rapidity, transverse momenta
are somewhat smaller, so that we overestimate $\Sigma p_T^2$. 
The spectrum (\ref{mtspectrum}) yields an average $\langle p_T^2 \rangle$ 
given by
\begin{equation}
\langle p_T^2 \rangle = 
 2m T \frac{1 + 3\frac{T}{m} + 3\frac{T^2}{m^2}}{1 + \frac{T}{m}}.
\label{pt2_1}
\end{equation}

Substituting in Eq.(\ref{pt2_1}) the values of the inverse slope 
parameter $T$ given above, and multiplying the resulting average 
squared transverse momentum by the multiplicity of the particle 
species, one obtains for a central collision at SPS.
\begin{equation}
\sum_j \langle p_T^2 \rangle \sim 930~\mbox{GeV}^2.
\label{pt2_val}
\end{equation}
As mentioned above, this is somewhat overestimated since we 
have neglected the decrease of inverse slopes away from 
mid-rapidity. 

For a non-central collision, we assume that the sum of $p_T^2$ 
scales with the multiplicity, i.e. we neglect the centrality 
dependence of inverse slopes and particle ratios. Therefore, in the 
case of semi-central collisions used by NA49 for flow studies, with 
only 40 to 55\% of the maximum multiplicity~\cite{Data}, the value
(\ref{pt2_val}) 
must be divided by 2.

\end{document}